# Molecular beam epitaxy of InAs nanowires in SiO₂ nanotube templates: challenges and prospects for integration of III-Vs on Si

Jelena Vukajlovic-Plestina[1], Vladimir G. Dubrovskii[2-4], Gözde Tütüncuoğlu[1], Heidi Potts[1], Ruben Ricca[1], Frank Meyer[1], Federico Matteini[1], Jean-Baptiste Leran[1], Anna Fontcuberta i Morral[1]

[1]Laboratory of Semiconductor Materials, Institute of Materials, Ecole Polytechnique Fédérale de Lausanne (EPFL), 1015 Lausanne, Switzerland

[2]St. Petersburg Academic University, Khlopina 8/3, 194021 St. Petersburg, Russia

[3]Ioffe Physical Technical Institute of the Russian Academy of Sciences, Politekhnicheskaya 26, 194021 St. Petersburg, Russia

[4]ITMO University, Kronverkskiy pr. 49, 197101 St. Petersburg, Russia

## Abstract

Guided growth of semiconductor nanowires in nanotube templates has been considered as a potential platform for reproducible integration of III-Vs on silicon or other mismatched substrates. Herein, we report on the challenges and prospects of molecular beam epitaxy of InAs nanowires on SiO₂/Si nanotube templates. We show how and under which conditions the nanowire growth is initiated by In-assisted vapor-liquid-solid growth enabled by the local conditions inside the nanotube template. The conditions for high yield of vertical nanowires are investigated in terms of the nanotube depth, diameter and V/III flux ratios. We present a model that further substantiates our findings. This work opens new perspectives for monolithic integration of III-Vs on the silicon platform enabling new applications in the electronics, optoelectronics and energy harvesting arena.

## 1. Introduction

One-dimensional geometry of semiconductor nanowires gives rise to many interesting physical properties which are not seen in bulk materials, such as electron quantum confinement [1–5] and optical resonances [6,7]. Furthermore, small diameters of NWs allow for their dislocation-free growth on lattice-mismatched substrates. This feature may enable monolithic integration of high-quality III–V semiconductors on to cost effective silicon substrate, overcoming the fundamental issues such as lattice, polarity and thermal expansion mismatch [8,9] To fully exploit the huge potential provided by the combination of NW geometry and excellent III-V semiconductors properties (direct band gap, superior carrier mobility and high absorption coefficient), the growth of these structures need to be achieved in Au-free and position controlled way. In this way any possible Au contamination of the nanowire or substrate is safely avoided [10–12]. In gold-free growth, the NW position is usually controlled by defining the arrays of holes in a dielectric mask using electron beam photolithography for surface patterning [13–15]. This provides the nanoscale precision, but on the other hand the process is extremely time-consuming on a large (wafer) scale and unable to satisfy the condition of cost effectiveness. Other techniques such as nanoimprint lithography [12,16,17] and phase shift lithography (PSL) [18] have previously been used to create large scale patterns for ordered NW growth.

In this work, we employ PSL as a patterning technique [18–20] to fabricate large areas of $SiO_2$ nanotubes on Si substrates to guide the subsequent growth of InAs nanowires by molecular beam epitaxy (MBE). A similar concept has recently been demonstrated by Borg *et al*. with metal-organic vapor phase epitaxy (MOVPE). This MOCVD technique is usually referred to as template-assisted selective epitaxy (TASE) [21,22]. MBE is characterized by highly directional material influxes, which makes it fundamentally different from MOVPE. The MBE template growth should be much more challenging since the material supply to the bottom of the nanotubes strongly depends on their depth. Consequently, we conduct a detailed experimental study of the template geometry influencing the yield of InAs nanowires, followed by a supporting theoretical model. Our results indicate that growth inside the nanotubes proceeds in the In-assisted vapor-liquid-solid mode rather than by selective area epitaxy. Furthermore, we demonstrate the nanowire growth in the tubes with a high aspect ratio, where the direct impingement onto the bottom of the tubes is no longer possible and the group III growth species are supplied only through surface diffusion and re-emission.

## 2. Experimental details

Large scale arrays of $SiO_2$ nanotubes were fabricated on (111)Si substrates. The fabrication process is shown in Fig. 1 a). First, Si nanopillars were defined by phase shift photolithography (PSL) followed by reactive ion etching, as in Refs. [18,23]. The pillars were ~500 nm high and with diameters ranging from 150 to 450 nm. A 50 nm thick layer of $SiO_2$ was grown around them by thermal oxidation. Further processing steps shown in Fig. 1 a) comprised coating with protective photoresist layer, controlled etching of photoresist up to the desired pillar height, oxide etching using HF and finally using reactive ion etching (RIE) to empty the silicon inside the nanopillars and form a $SiO_2$ nanotube. The entire process was optimized for 4 inch wafers, which were diced in 4 substrates for growth in our molecular beam epitaxy machine. Each growth substrate contained 5 arrays with different geometries, with pitches of 1, 1.5 and 2 µm pillars and diameters varying from 150 nm up to 450 nm. The depth of the tubes varied between 50 and 400 nm. A scanning electron microscopy (SEM) image of a typical oxide nanotube template (ONT) array is shown in Fig. 1 b). The arrays and the nanotube morphology appear very homogenous across the wafer.

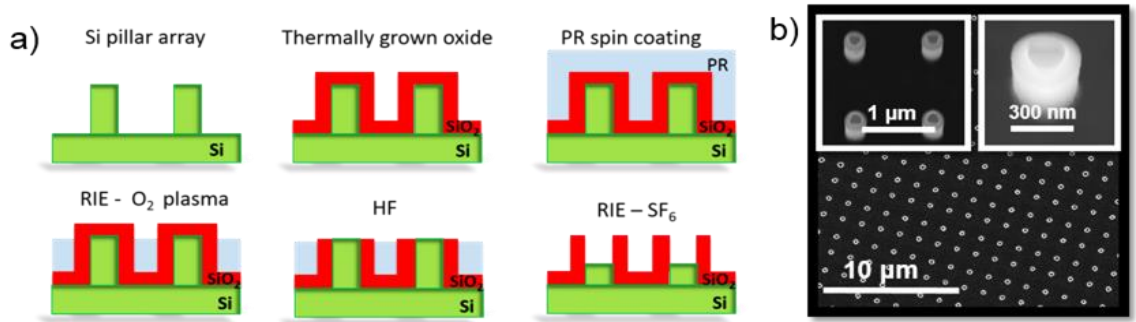

**Figure 1** a) Illustration of a silicon nanotube (ONT) array fabrication steps and b) SEM images of an ONT array and close-view of a single nanotube.

InAs nanowires were then grown in the ONTs in a DCA P600 MBE system. Before introduction into the MBE growth chamber, samples were dipped for 2 s in poly-silicon etch solution [$HNO_3$(70%):HF(49%):$H_2O$] in order to remove the native oxide and smoothen the silicon surface [24]. Growth parameters (the substrate temperature $T_S$, the growth time $t$, the In and $As_4$ beam equivalent pressures $P_{In}$ and $P_{As}$) were systematically varied one parameter at the time. The optimal growth temperature was found to be 500˚C, regardless of the nanotube geometry. At this temperature, we obtain growth inside the nanotube templates and at the same time avoid parasitic growth outside the nanotubes. The In and $As_4$ beam equivalent pressures were varied between 1.1 and 1.8x$10^{-7}$ Torr and 0.7 and 1.3x$10^{-5}$ Torr, respectively, similar to the conditions used in prior works for InAs nanowire growth [7]. Growth time was varied between 30 min and 5h. No dependence on the array pitch was observed.

## 3. Results and discussion

### 3.1. Optimizing material supply for InAs nanowire growth: 200 nm deep tubes

We start by recounting how the nanotube aspect ratio affects the material supply onto the bottom of the tube. A sketch of the nanotube geometry is shown in Fig. 2 a). As mentioned above, the As and In beams are directional, In adatoms are able to diffuse on the surface while As is highly volatile and almost non-diffusive [25-27]. The depth $H$ and diameter $D$ of the nanotube determine the amount of material reaching its bottom to start the nanowire growth. In our MBE system, both cells are positioned at 45°. Hence, at $H > D$ there is no direct impingement of both species at the bottom of the nanotube. In this case, initiation of the nanowire growth should be an extremely slow process as only diffusing species can reach the silicon bottom of the nanotube. Diffusion lengths of In on $SiO_2$ are typically between 0.5 and 0.8 µm, depending on the growth conditions [7,25,15]. The arsenic species ($As_2$ and $As_4$) should have the diffusion lengths of a few nm at most, however, arsenic can be re-emitted from the ONT surfaces and thereby contribute to the InAs growth even in the shadowed areas [26,27]. In fact, the ONT creates local growth conditions inside the nanotubes which can be significantly different from the nominal conditions for two-dimensional (2D) growth. Therefore, in Fig. 2 a) the effective In (labeled "3") and As (labeled "5") atomic fluxes that reach the bottom of the nanotube are denoted $\chi_3 v_3$ and $\chi_5 v_5$, with $\chi_k$ containing information on the geometry and re-emission of the growth species within the nanotube.

We first investigated growth in the nanotubes whose diameters are similar to the depth ( $D \simeq H$, aspect ratio $H/D \simeq 1$) using the growth conditions that yield InAs NW arrays on patterned Si substrates ($T_S$ = 500°C, $P_{In}$ = 1.2x10$^{-7}$ Torr, $P_{As}$ = 6x10$^{-6}$ Torr) and the growth time of 60 min [5]. These conditions did not produce any nanowires inside the nanotubes, supporting that the effective V/III flux ratio inside the nanotube is different from the nominal value. As the diffusion length of In is much larger than that of As, much more In is collected inside the nanotube and the actual V/III ratio might be too low for nanowire growth. We gradually increased the growth time, the V/III ratio as well as the absolute fluxes of both As and In fluxes. Figures 2 b) to d) show the effect of increasing the $P_{As}$ value at the fixed $T_S$ = 500°C, $P_{In}$ = 1.4x10$^{-7}$ Torr and 90 min growth time. Below the SEM images, we show the corresponding yields measured for all types of structures: vertical nanowires, non-vertical nanowires, nanoscale V-shaped membranes [28], quasi-2D parasitic growth and empty nanotubes (i.e., absence of any growth). It is important to note that in many cases non-vertical nanowires and nanoscale V-shaped membranes nucleate on the oxide layer rather than on Si surface inside the nanotubes [see Supporting Information (SI) 1].

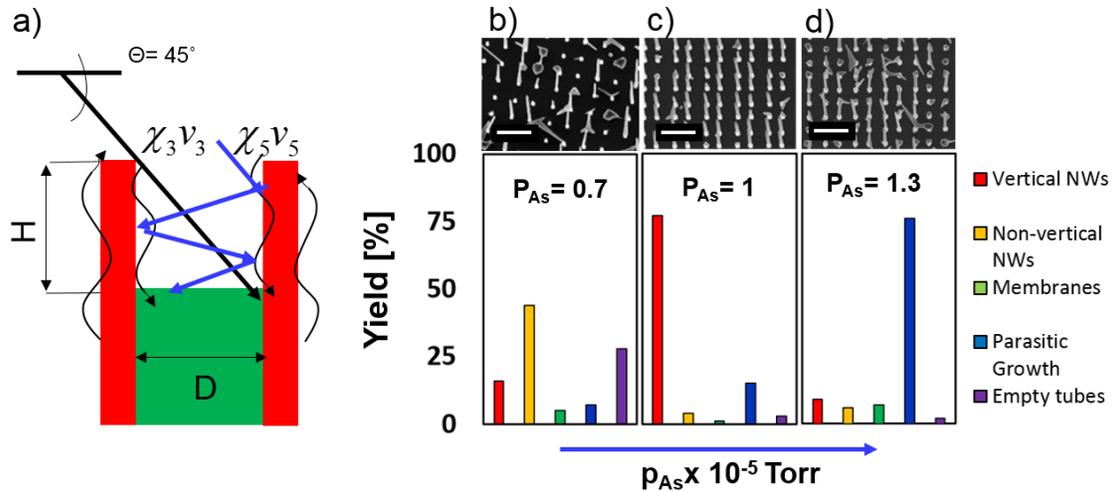

**Figure 2** a) Sketch of the SiO$_2$/Si nanotube with the effective In and As fluxes $\chi_3 v_3$ and $\chi_5 v_5$, respectively. b) to d) Arsenic series of the InAs nanowire growth in the ONT with the resulting morphology presented top right. All samples were grown at $T_S$ = 500°C and $P_{In}$ = 1.4x10$^{-7}$ Torr for 90 minutes. The histograms show the yields of different structures obtained from the statistics analysis of the SEM images. The scale bar for the SEM images is 2 μm and the tilt angle is 20°. The $P_{As}$ values are given in the units of 10$^{-5}$ Torr.

For the lowest values of $P_{As}$ around 0.7x10$^{-5}$ Torr we observe a very low yield of vertical nanowires (less than 20%). Many non-vertical nanowires, some membranes and parasitic structures nucleate on the surface and many tubes remain empty. One can conclude that this $P_{As}$ is insufficient to nucleate vertical nanowires in the desired nucleation position, i.e. on bare Si surface in the bottom of the nanotubes. Increasing $P_{As}$ to 1x10$^{-5}$ Torr leads to a significant increase in the yield of vertical NWs up to 77%. The unwanted non–vertical nanowires, membranes are greatly suppressed (down to 8% in total), in 10% of the tubes the parasitic growth was found and very few empty tubes remain (5%). By further increasing $P_{As}$ to 1.3x10$^{-5}$ Torr, we again obtain a dramatic change. In this case, less than 5% of empty tubes are observed, while vertical nanowires (~10%) are replaced by parasitic structures (almost 75%), non-vertical nanowires and membranes (about 10% in total). These noticeable differences in the yields of vertical NWs versus other structures are clearly visible on the SEM images shown in Figs. 2 b) to d). Qualitatively, parasitic growth could be due to the reduction of the In diffusivity for higher As fluxes [29,30].

Table 1. Comparison between the optimal growth conditions for InAs nanowires on standard patterned Si substrates and ONTs with a tube depth of 200 nm and aspect ratio of one

| Substrate type | $T_s$ [°C] | $p_{In}$ x 10$^{-7}$ [Torr] | $p_{As}$ x10$^{-5}$ [Torr] | t [min] |
|---|---|---|---|---|
| Standard patterned Si substrate [5] | 500 | 1.2 | 0.6 | 60 |
| Nanotube templates; depth 200 nm | 500 | 1.4 | 1 | 90 |

Figure 3 shows the representative SEM images of the sample with the highest yield of vertical nanowires after 90 min of growth. The high yield is homogeneous over the area of the array. In Table 1, we compare the optimized growth conditions for the high yield of vertical InAs nanowires on standard patterned Si substrates [5] and in the 200 nm nanotubes with $D \simeq H$. Clearly, higher fluxes and V/III ratios are required to grow nanowires in the ONTs.

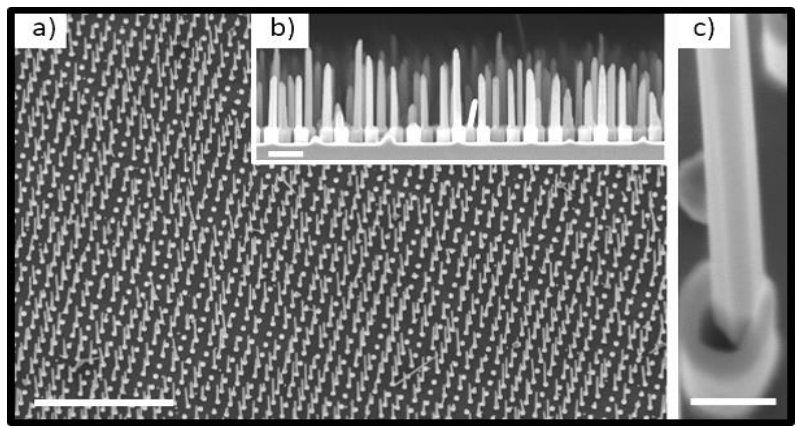

**Figure 3** a) InAs nanowires grown under optimized conditions in the ONTs with the tube depth of 200 nm and the aspect ratio of one. The scale bar is 10 µm and the tilt angle is 20°. b) Cross section of the same sample. The scale bar is 1 µm. c) Single NW growing from the nanotube, the scale bar is 200 nm and the tilt angle is 20°.

In order to understand the growth mechanism, we now take a closer look at the nanowires growing in the nanotubes. Typical SEM image of an InAs nanowire growing vertically from a nanotube is shown in Fig. 3 c). It is clearly seen that the nanowire does not fully fill the nanotube volume. Therefore, InAs nanowires do not start growing on the entire available area of bare Si in the bottom of the nanotube.

This strongly suggests that the growth inside the nanotube does not proceed via the selective area epitaxy mode. Rather, at a low effective V/III ratio inside the nanotube (compared to its nominal value for 2D growth), locally In-rich conditions are very favorable for nucleation of In droplets that can subsequently promote the In-catalyzed vapor-liquid-growth of InAs nanowires [31,32]. A similar reasoning was recently reported by Robson *et al*. for the growth of InAs nanowires on patterned Si substrates where the initial nucleation step was In-assisted [33]. The nanowires should then be positioned at the edges of the nanotubes because In droplets have better chances to nucleate at the tube corners for surface energetic reasons (i.e., replacing the energetically costly liquid-vapor surface to less energetic liquid-solid interface [34,35]).

We now analyze the influence of the nanotube diameter on the vertical nanowire yield, keeping the nanotube depth at 200 nm. The representative SEM images are shown in Fig. 4. For 160 nm diameter, we observe a large variety of structures: non-vertical wires, membranes and empty tubes [Fig. 4 a)]. For diameters between 200 and 350 nm, uniform vertical nanowires are obtained [Fig. 4 b) and c)]. Interestingly, InAs nanowires stop growing for diameters larger than 500 nm [Fig. 4 d)]. These results further support the idea of local growth conditions created inside the nanotubes. They also show that the different template openings will require different growth conditions to produce vertical nanowires.

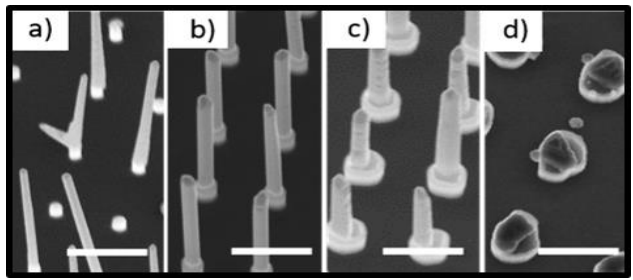

**Figure 4.** SEM images of InAs NWs grown in ONTs. The tube depth $H$=200 nm and the growth conditions were the same for all samples. The only parameter varied was the tube diameter $D$: a) 160 nm, b) 200 nm, c) 350 nm and d) 550 nm. The scale bar is 1 µm and the tilt angle is 20° for all images.

### 3.2. Effect of nanotube depth

The next parameter explored was the nanotube depth $H$. Clearly, the $H$ (or $H/D$) value strongly influences the local V/III ratio inside the ONTs, with the case of $H>D$ resulting in a very poor supply of the growth species onto the bottom of the nanotubes. This should lead to a longer delay before the nanowire growth can start. Figure 5 shows the nanowire arrays obtained after 90 min of growth with the parameters optimized for 200 nm deep tubes ($T_S$ = 500°C, $P_{In}$ = 2x10$^{-7}$ Torr, $P_{As}$ = 1x10$^{-5}$ 90 minutes) and variable $H$ =

50, 200 and 400 nm. The exact tube depths were determined from the cross-sections prepared by ion beam thinning, as shown in the insets of Fig. 5. The yield of vertical nanowires in shallow nanotubes (50 nm) is about 70 %; however, we observe more V-shaped membranes than in 200 nm deep tubes. In this particular case, most membranes nucleate on Si surface inside the nanotubes rather than on the oxide surface (see SI 2). Since smaller depths should relate to higher arsenic inputs, this results is in agreement with Ref [28] where higher V/III ratios gave higher yield of membranes.

As expected, increase of the depth to 400 nm leads to the growth of very few structures and instead most nanotubes remain empty. The absence of growth in this case could be due to insufficient time to nucleate the structures under a low material supply and probably inappropriate effective V/III ratio in the bottom of the nanotubes under these conditions. Therefore, below we present a more detailed growth study in 400 nm deep nanotubes.

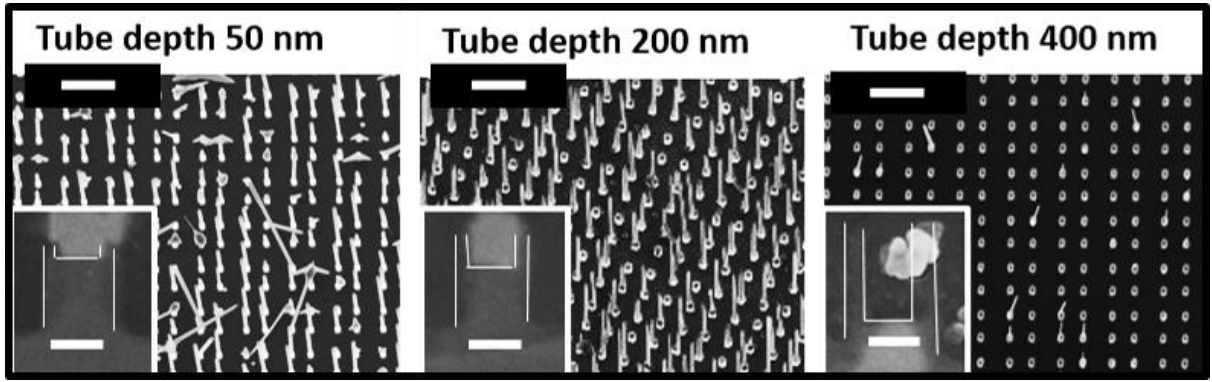

**Figure 5** InAs NWs growth in the ONTs with different tube depth. The scale bar in the SEM images is 1 μm and the tilt angle is 20˚. The scale bars in the cross-sectional images is 200 nm.

### 3.3. Growth in deep nanotubes

Considering that the growth in nanotubes with $H>D$ is controlled by surface diffusion and re-emission and hence should be much slower than in shallow tubes, we have first explored the effect of growth time under the same growth conditions as in sections 3.2 and 3.3. The representative SEM images for this time series are shown in Fig. 6. Clearly, increasing the growth time has a positive effect on the yield. After 1.5 hours of growth, very few wires were obtained and tubes mainly remained empty, as presented in the [Fig.6 a)]. The insert in the same figure show closer view to the single NW growth from the tube (left) and empty tubes (right). For the 5 hours growth the yield is pointedly improved [Fig.6 c)]. The measured overall yield of all structures (mainly vertical and tilted wires) is about 2%, 4% and 67% for the 1.5, 3 and 5 hours growths, respectively. A closer look at the 5 hours growth results reveals a remarkable difference in the nanowire morphologies that co-exist in one sample. One can observe thin and single crystalline nanowires, similar to those grown for shorter times, and nanowires with multigrain structure whose crustal quality has significantly degraded with respect to the shorter growth times. These differences are shown in the inset to Fig 6 c).

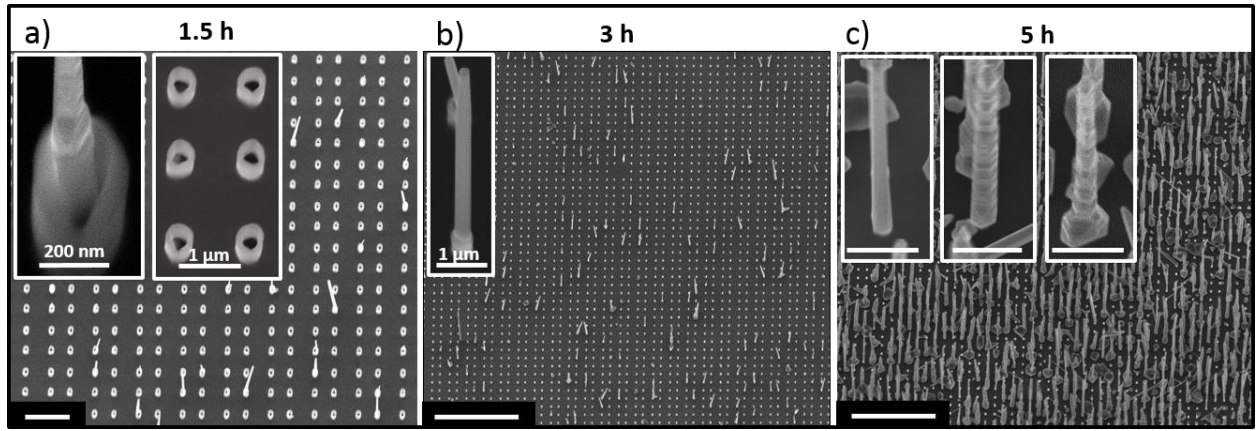

**Figure 6**. 20° tilted SEM images illustrating the time evolution of the morphology in 400 nm deep ONTs. The scale bare is 2 µm in the 1.5 hours image and 10 µm otherwise.

The images shown in the insert to Fig. 6 a) reveals that some nanowires exhibit different morphologies inside and outside the nanotube e.g. the diameter shrinks outside of the tube and the crystal structure becomes visibly more defective. This is consistent with the local environments for growth being different in the initial nucleation stage and after the nanowire leaves the template. As in section 3.1, the nanowires are positioned at the nanotube edges. We have also tried to improve the vertical yield in 400 nm deep nanotubes by increasing the material fluxes and keeping the growth time at 90 min. However, this just led to an increase in parasitic growth. The details od this study are given in SI 3.

### 3.4. Theoretical model

We now turn to physical modeling of MBE growth of InAs nanowires in ONTs. We will try to explain the experimentally observed trends such as

(i) The optimum arsenic flux to obtain high vertical yield should be neither too low nor too high;

(ii) It is more difficult to grow regular nanowires in deep tubes (with $H > D$);

(iii) There is an optimum tube diameter range for a given tube height giving the highest vertical yield (for example, $D=200$-350 nm for $H=200$ nm).

As in Ref. [33], we assume that high vertical yields are achieved by the mononuclear vapor-liquid solid growth in the initial stage, assisted by In droplet as illustrated by scenario (I) in Fig. 7 a). This view is supported by the fact that most vertical nanowires do not fully cover the template bottom. Scenario (II) in Fig. 7 a) corresponds to strongly As-rich conditions, leading to the true selective area epitaxy. The signature of this growth mode would be the completely filled template, which was not observed [one should not mix this case with the template filled by the radial nanowire growth in a later stage, as seen in Figs. 6 b) and c)]. On the other hand, in scenario (III) with the excessive In influx, the droplet will inflate too quickly and the nanotube will soon be filled with In liquid. This liquid will subsequently spread out of the tube, producing multiple and irregular structures. Scenario (IV) in Fig. 7 a) illustrates the polynuclear growth regime [36] in which two or more droplets emerge in one tube, enabling the growth of more than one nanowires per tube. This is not desirable for growing regular and single-crystalline nanowires, because radial merging of neighboring nanowires often lead to the formation of poly-crystallites [37].

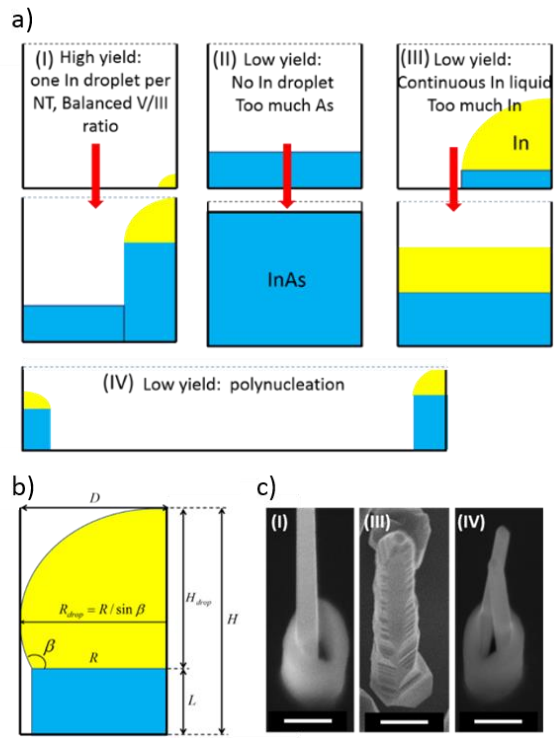

***Figure 7*** *(a) Cross-sectional sketches of the nanotubes. (I) High vertical yield under a balanced V/III ratio resulting in mononuclear In-assisted nucleation of nanowires; (II) Strongly As-rich conditions corresponding to the selective area growth; (III) Strongly In-rich conditions producing continuous In liquid within the tube; (IV) Polynucleation resulting in irregular poly-crystallites, (b) Model geometry showing the maximum*

*droplet size within the nanotube template; c) 20° tilt SEM images supporting scenarios (I), (III) and (IV). The scale bare is 200 nm for all SEM images.*

Continuous In liquid starts forming in the tube in the limiting geometry shown in Fig. 7 b). The nanowire length including the droplet reaches the tube height when $L + H_{drop} = H$, where $L$ is the nanowire length. Assuming that the nanowire is half a cylinder, the arsenic-limited regime of axial nanowire growth [38-41] yields the linear time dependence $L = 2\chi_5 v_5 t$. If we assume that all In atoms arriving into the tube at the rate $\chi_3 v_3$ are subsequently collected by the droplet whose contact angle $\beta$ remains time-independent, the radius of the droplet base $R$ will grow with time as in Ref. [38]: $R = [2\Omega_3 / \Omega_{35} f(\beta)](\chi_3 v_3 - \chi_5 v_5) t$. Here, $\Omega_3$ is the elementary volume in the In liquid, $\Omega_{35}$ is the volume of InAs pair in the solid state and $f(\beta)$ is the geometrical function relating the volume of half a spherical cap to the radius of its base. Now, the droplet width $R_{drop}(t_*) = R(t_*) / \sin \beta$ in scenario (1) must remain smaller than the tube diameter $D$ by the moment of time $t_*$ at which $L(t_*) + R(t_*)(1 - \cos \beta) / \sin \beta = H$. Using the above equations for $L(t)$ and $R(t)$, this condition is quantified as

$$\frac{\chi_5 v_5}{\chi_3 v_3} > \frac{c_3 [H - D(1 - \cos \beta)]}{D + c_3 [H - D(1 - \cos \beta)]} \quad , \quad (1)$$

where $c_3 = \Omega_3 / [\Omega_{35} f(\beta) \sin \beta]$ is the shape constant. The $c_3$ value equals 0.27 for the contact angle $\beta$ of 120°.

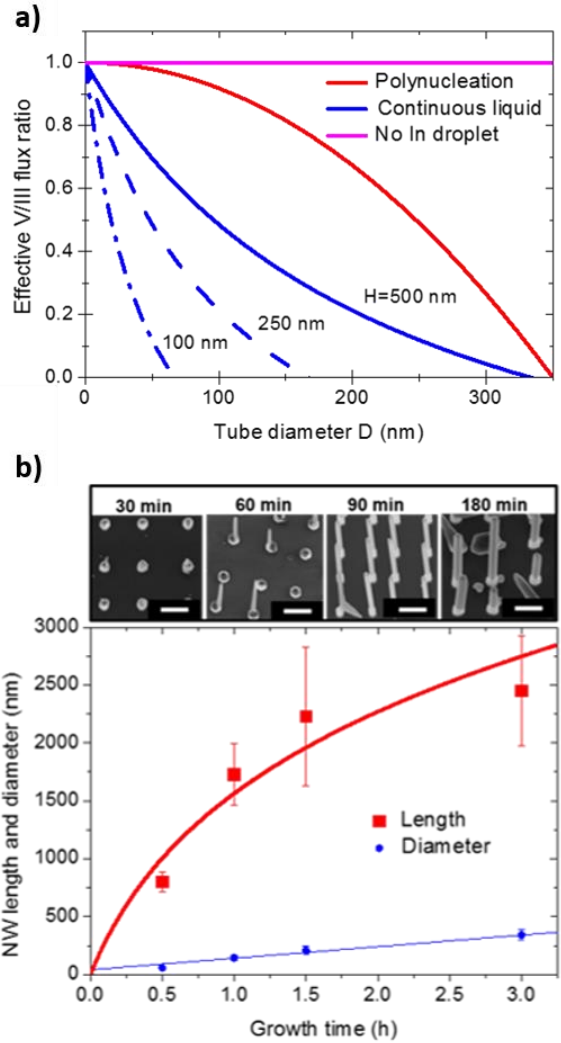

**Figure 8** a) Diagram showing the optimum regions for high yield of vertical InAs nanowires in the ONTs in in terms of the effective V/III flux ratio and the nanotube diameter. The best zone is separated by the upper limiting curve corresponding to polynucleation in the tubes and the lower limiting curve corresponding to the formation of continuous indium liquid in the tubes. The V/III ratio of one corresponds to the absence of the droplet nucleation or their consumption, b) Mean nanowire radius and length versus time: experimental data (symbols) fitted by the model (lines) with the corresponding 20° tilted SEM images shown above. The nanowires were grown in the ONTs with nominal depths 50-100 nm. The growth conditions used were $T_s$ = 500°C, $P_{In}$ = 1.4x10$^{-7}$ Torr, $P_{As}$ = 1x10$^{-5}$. The scale bare is 1 µm.

On the other hand, the mononuclear growth within the tubes requires that the waiting time between two successive nucleation events for In droplets, $t_{nucl} = 1(\pi D r_c J_{corner})$, is longer than the time $t_{growth}$ required to fill the template bottom by one nanowire base [36]. Here, $r_c$ is the radius of the critical nucleus and $J_{corner}$ is the nucleation rate at the corner of the tube. The $t_{growth}$ can be approximated as $t_{growth} \cong D/[2c_3(\chi_3 v_3 - \chi_5 v_5)]$. This yields a lower limit for the tube diameter of the form $D^2 < [2c_3(\chi_3 v_3 - \chi_5 v_5)]/[\pi r_c J_{corner}]$. Combining this with Eq. (1), we obtain the two conditions for high vertical yield

$$1 - \frac{D^2}{D_{nucl}^2} > \frac{\chi_5 v_5}{\chi_3 v_3} > \frac{c_3[H - D(1-\cos\beta)]}{D + c_3[H - D(1-\cos\beta)]} \qquad (2).$$

Here, the characteristic "nucleation" diameter is given by $D_{nucl}^2 = (2c_3 \chi_3 v_3)/(\pi r_c J_{corner})$ and increases with the In flux $v_3$. Therefore, for a given geometry, there is an optimal range of the effective V/III flux ratios to avoid both polynucleation (lower limit) and overloading the template with liquid In (upper limit). The SEM images shown in Fig. 7 c) perfectly support the existence of scenarios (I), (III) and (IV).

Figure 8 a) shows the corresponding diagrams for the typical $\beta = 120°$ [38–40], $c_3 = 0.27$ and $D_{nucl} = 350$ nm, at three different tube depths $H$. These graphs explain qualitatively the major effects. First, the vapor-liquid-solid nucleation of InAs NWs is more difficult in deeper tubes. In fact, increasing $H$ can reduce the optimum region in Fig 8 a) to nothing. Second, for a given $H$, there are the optimum regions in both the nanotube diameters $D$ and effective V/III flux ratios $\chi_5 v_5 / \chi_3 v_3$ to grow nanowires with high yields, as observed experimentally (Sections 3.1 and 3.2).

Up to now, we were focusing on the initial growth stage which proceeds as long as the nanowire is by its full length within the template. After the nanowire leaves the template, the In collection becomes less effective and the As flux onto the droplet increases, both effects leading to increasing the actual V/III influx ratio into the droplet. According to the diagram shown in Fig. 8 a), this should reduce the droplet size until it disappears completely, as in Refs. [33,39]. After that, the growth is transitioned to the vapor-solid mode and becomes limited by the material transport of In atoms to the nanowire top [42,43]. Consistent with our experimental observations, we assume that the nanowire radius continues increasing linearly with time due to the In incorporation on the sidewalls:

$$R = R_0 + v_R t. \qquad (3)$$

Here, $R_0$ is the initial nanowire radius at the beginning of this growth stage and $v_R$ is the radial growth rate. The axial elongation can be written in the form [42]

$$\frac{dL}{dt} = v_3\left(1 + \frac{2\varphi_3}{\pi}\frac{\lambda_3}{R}\right), \qquad (4)$$

with $\varphi_3$ as the indium collection efficiency at the nanowire sidewalls and $\lambda_3$ as the diffusion length of In adatoms.

Using Eq. (3) in (4) and integrating, we obtain

$$L = v_3 t + \Lambda_3 \ln\left(1 + \frac{v_R t}{R_0}\right), \qquad (5)$$

with $\Lambda_3 = (2\varphi_3 v_3 / \pi v_R)\lambda_3$ as the effective collection length of indium on the top part of the NW sidewalls. The unusual logarithmic dependence arises due to lateral growth, and is converged to the more common expression $dL/dt = v_3 t[1 + (2\varphi_3\lambda_3)/(\pi R)]$ only for small times. For long enough growth times, the nanowires elongate at a lower rate according to $L \cong v_3 t$, as observed in our experiments. Figure 8 b) shows the reasonable fits by Eqs. (3) and (5) to the measured time dependences of the mean length and radius, obtained with $R_0 = 20$ nm, $v_R = 50$ nm/h, $v_3 = 30$ nm/h and $\Lambda_3 = 1200$ nm. The representative SEM images of the corresponding time series are shown in the inserts to Fig. 8 b). More details on the radial growth are given in SI 4.

## 4. Conclusion

We have demonstrated the MBE template growth of InAs nanowires in large scale silicon dioxide nanotubes on silicon. Geometrical parameters such as the nanotube depth and diameter have been investigated in order to maximize the vertical nanowire yield. The most critical parameter for such growth is the As flux or the effective V/III flux ratio. It has been shown that the maximum vertical yield is achieved for a balanced V/III ratio which should be neither too high nor too small for a given geometry of the ONT. We have presented evidences of In-assisted vapor-liquid-solid growth in the initial stage within the nanotubes under local conditions that are different from the vapor environment. Our theoretical model explains satisfactorily the relation between the growth conditions and the nanotube geometry for obtaining the high vertical yield, as well as the nanowire growth kinetics in a later stage. Overall, this study constitutes the first step toward using the $SiO_2$ nanotubes as templates for the cost-effective and Au-free MBE growth of III-V nanowires on large area silicon substrates.

**Acknowledgement**

Authors thank the CIME and in particular D. Laub for the fabrication of cross-sections. V.G.D. acknowledges financial support received from the Ministry of Education and Science of Russian Federation under grant 14.613.21.0055 (project ID: RFMEFI61316X0055). VGD, AFiM and JVP thank thank FP7 funding through the project Nanoembrace. AFM thanks funding of the Eranet program of SNF through project nr IZLRZ2_163861, the NCCR QSIT, the ERC Starting Grant UpCon as well as H2020 thorugh project INDEEd.